\documentclass[twocolumn]{svjour3}
\usepackage[dvipsnames]{color}
\usepackage{mathtools,cuted}
\usepackage{amsmath}
\usepackage{graphicx}
\usepackage{latexsym}
\usepackage{mathrsfs}
\usepackage{amssymb,amsbsy}
\usepackage{gensymb}
\usepackage{amsfonts}
\usepackage{epstopdf}
\usepackage{multirow}
\usepackage{array}
\usepackage[caption=false,font=footnotesize]{subfig}

\def\be{\begin{equation}}
\def\ee{\end{equation}}
\renewcommand{\deg}{^{\circ}}
\newcommand{\as}{^{\prime\prime}}
\newcommand{\rss}{\scriptscriptstyle}

\begin{document} 

\title{A software package for evaluating the performance of a star sensor operation}

\author{Mayuresh Sarpotdar \and Joice Mathew \and A.G. Sreejith \and Nirmal K. \and S. Ambily \and Ajin Prakash \and Margarita Safonova \and Jayant Murthy}
\institute{Indian Institute of Astrophysics, Bangalore, India.\\
\email{mayuresh@iiap.res.in} }
\date{Received: 09 August 2016 / Accepted: 26 December 2016}
\maketitle

\begin{abstract} 
We have developed a low-cost off-the-shelf component star sensor ({\em StarSense}) for use in minisatellites and CubeSats to determine the attitude of a satellite in orbit. {\em StarSense} is an imaging camera with a limiting magnitude of 6.5, which extracts information from star patterns it records in the images. The star sensor implements a centroiding algorithm to find centroids of the stars in the image, a Geometric Voting algorithm for star pattern identification, and a QUEST algorithm for attitude quaternion calculation. Here, we describe the software package to evaluate the performance of these algorithms as a star sensor single operating system. We simulate the ideal case where sky background and instrument errors are omitted, and a more realistic case where noise and camera parameters are added to the simulated images. We evaluate such performance parameters of the algorithms as attitude accuracy, calculation time, required memory, star catalog size, sky coverage, etc., and estimate the errors introduced by each algorithm. This software package is written for use in MATLAB. The testing is parametrized for different hardware parameters, such as the focal length of the imaging setup, the field of view (FOV) of the camera, angle measurement accuracy, distortion effects, etc., and therefore, can be applied to evaluate the performance of such algorithms in any star sensor. For its hardware implementation on our {\em StarSense}, we are currently porting the codes in form of functions written in C. This is done keeping in view its easy implementation on any star sensor electronics hardware.


\end{abstract}
   

\section{Introduction}
\label{sec:intro}

An attitude/orientation control system is necessary for all satellites with different satellites requiring varying degrees of pointing accuracy. Satellites with a high-gain directional antenna, a telescope, or an Earth-imaging instrument as payloads require highly accurate (few arcminutes or arcseconds) pointing system. Determining the current pointing position is of primary importance in such applications and it is usually achieved using star sensors: a wide-FOV camera with online image processing capabilities which reduces the image data to an attitude quaternion that describes the rotation of the sensor coordinate system with respect to the Earth-centered inertial (ECI) coordinate system. This is achieved by applying multiple algorithms on the image in sequence \cite{survey_on_algorithms}. We have developed a low-cost star sensor {\em StarSense} to deploy on minisatellites and CubeSats using off-the-shelf components, in which we have implemented a centroiding algorithm for finding centroids of stars seen in the image, a Geometric Voting algorithm for star pattern identification, and a Quaternion Estimator (QUEST) algorithm for quaternion estimation. To determine how each algorithm performs in terms of sky coverage, memory requirements, calculation time, and so on, we simulate the algorithms along with idealized and real hardware parameter inputs using a specially developed testing software package. 

In this paper, we present a complete software package intended to evaluate the performance of algorithms implemented on the {\em StarSense}. This software package can be applied to any star sensor by changing the hardware parameters, such as the focal length of the imaging setup, the field of view (FOV) of the camera, angle measurement accuracy, distortion effects, and others. It is written to run under MATLAB due to the simplicity of scripting and its excellent capability of visualizing the results. The implementation on actual star sensor electronics hardware will be done in C, keeping in view its easy portability to other platforms. 

In Section~\ref{sec:hardware_info}, we briefly describe the hardware of the star sensor. Section~\ref{sec:software} describes the architecture of the software package we used to evaluate the algorithms. Section~\ref{sec:algorithms} describes the algorithms in detail. In Section~\ref{sec:performance_est}, we  describe the testing methodology to analyze the performance parameters of these algorithms, and the final results from different tests.

\section{Hardware Implementation}
\label{sec:hardware_info}

{\em StarSense} is essentially a wide-field imaging camera with high sensitivity. The optics consists of a four-element Tessar lens system  designed to minimize chromatic aberration, coma, and distortion. The lens system is designed with weight as a limiting constraint, and a condition of seeing a minimum of 3 stars in FOV in any field of the sky. The lens assembly is designed to sustain vibrations experienced during the satellite launch. A baffle is used to prevent the stray-light from the Sun and the Earth entering the optical system, thereby contributing to the sky background. In addition, we have used a thermal cut-down filter (a hot mirror), which reflects the IR and UV wavelengths and only allows optical wavelengths from 450--750 nm to pass through, to prevent the detector from getting heated by the direct Sun. The imaging sensor is the only electronic component directly open to the radiation environment and, therefore, its performance is prone to degradation with time. We have used a radiation-hardened CMOS detector Star-1000 \cite{Star1000}, which is sensitive to low light conditions. The optics has been designed specifically for this detector, so that the point spread function (PSF) is maintained at a full width at half maximum (FWHM) of 2 pixels, which corresponds to a star spread of $4\times4$ pixels, at all field positions.

\begin{figure}
\centering
\includegraphics[scale=0.08]{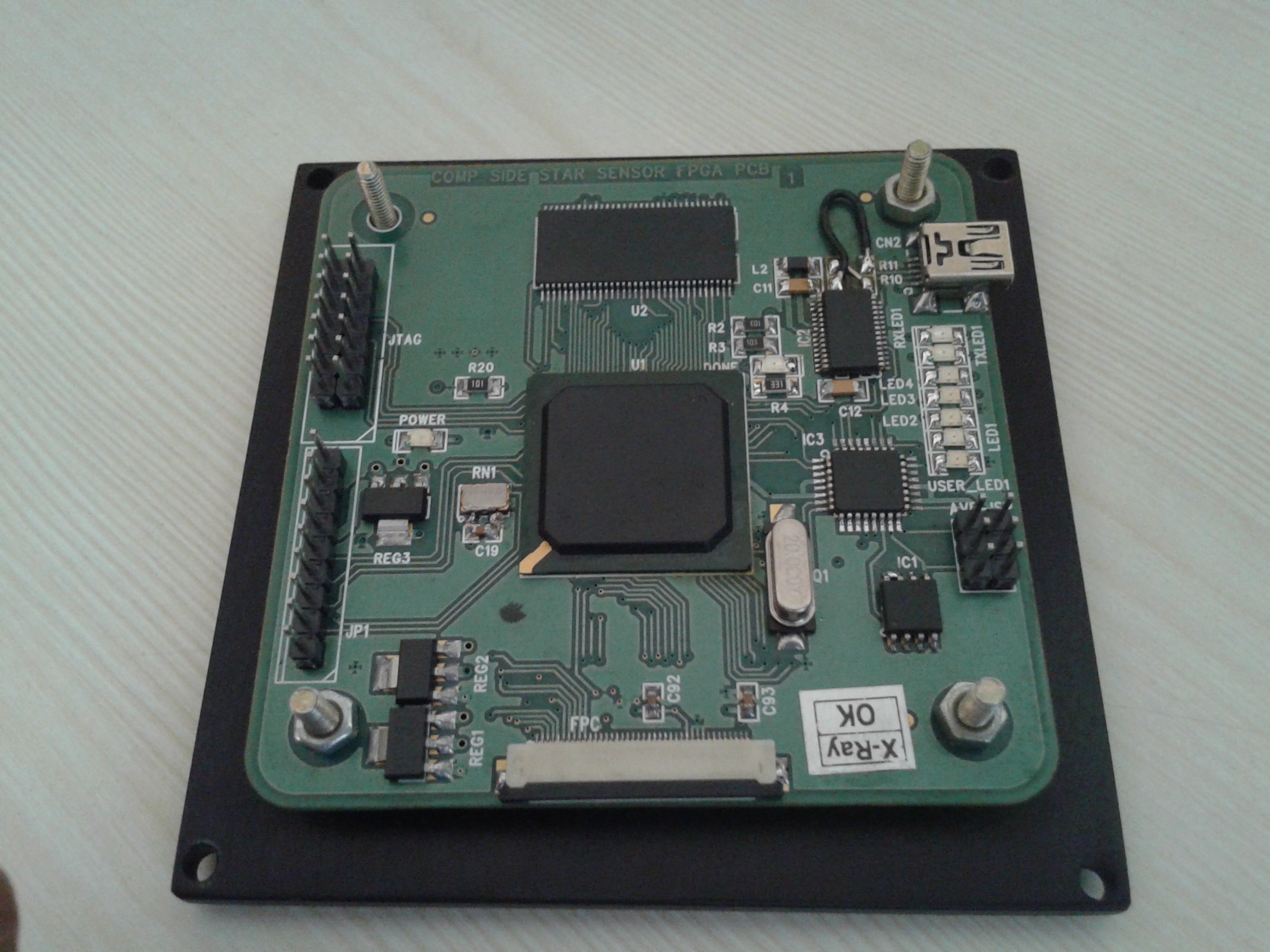} 
\caption{{\em StarSense} image processor board}
\label{fig:processor_board}
\end{figure}

The electronic system of \emph{StarSense} is inspired by Cubestar \cite{overall_paper_1}. It is distributed over two separate printed circuit boards (PCBs): detector PCB and image processor PCB. The image sensor and its biasing circuits are soldered on the detector PCB, which is fixed inside the star sensor structure in such a way that the focal plane of the star sensor optics coincides with the image sensor plane. The detector is connected to the image processor board through a flat plastic cable. Such an arrangement allows for the image processor board to be removed from the structure without disturbing the focal plane alignment of the detector PCB. The readout system for the image sensor is implemented using a MIL-Grade Spartan-6 field programmable gate array (FPGA) \cite{spartan 6 MIL grade} on the image processor PCB. The PCB also hosts SDRAM, required for the online processing of images, and a flash memory for nonvolatile storage of the bitstream and a star catalog. Figure~\ref{fig:processor_board} shows the {\em StarSense} image processor board. 

Since the optics, electronics and the structure were fully designed in-house, their characteristic properties and modeling parameters are well known. These parameters are used as inputs to algorithms discussed further in this paper. A lens with a field of view of $10\deg$, corresponding to a focal length of 80 mm for the selected detector, was chosen following a sky  simulation using Hipparcos bright star catalog. The key parameters of the camera are described in Table~\ref{table:camera_param}, and the key features of the electronics are presented in Table~\ref{table:star_sensor_electronics}. The camera parameters have been verified using different calibration and characterization tests of the complete star sensor assembly, such as dark signal, flat field, boresight alignment with respect to housing, field of view, distortion, magnitude calibration and linearity. A detailed paper justifying the choice of the hardware and various design parameters, including the results of calibration tests, is forthcoming. 

\begin{table}[h!]
\centering
\caption{{\em StarSense} Camera Parameters} 
\begin{tabular}{ll}
\hline
Field of View & $10\degree$ circular FOV\\
Focal Length & 80 mm\\
Aperture & F/2.6\\
Limiting Magnitude & V=6.5\\
Image size & $1024 \times 1024$\\
Pixel size & $15\times 15$ $\mu$m$^2$\\
Pixel scale & $36^{\prime\prime}$/px\\
Integration time & 100 ms\\
ADC resolution & 10 bit\\
Weight & 600 gm\\
Power& 2 W\\
\hline
\end{tabular}
\label{table:camera_param}
\end{table}

\begin{table}[h!]
\centering
\caption{{\em StarSense} Electronics Specifications}
\begin{tabular}{ll}
\hline
Detector & Star 1000 (Radiation-hardened)\\
Readout & MIL Grade Spartan-6 FPGA (XQ6SLX150T)\\
Non-volatile memory & 8 MB (For star catalog and bit stream)\\
RAM & 64 MB (For on-board image processing)\\
Operating Voltage & 5 V\\
\hline
\end{tabular}
\label{table:star_sensor_electronics}
\end{table}


\section{Software Implementation}
\label{sec:software}

\begin{figure}
\centering
\begin{tabular}{c}
\includegraphics[scale = 0.4]{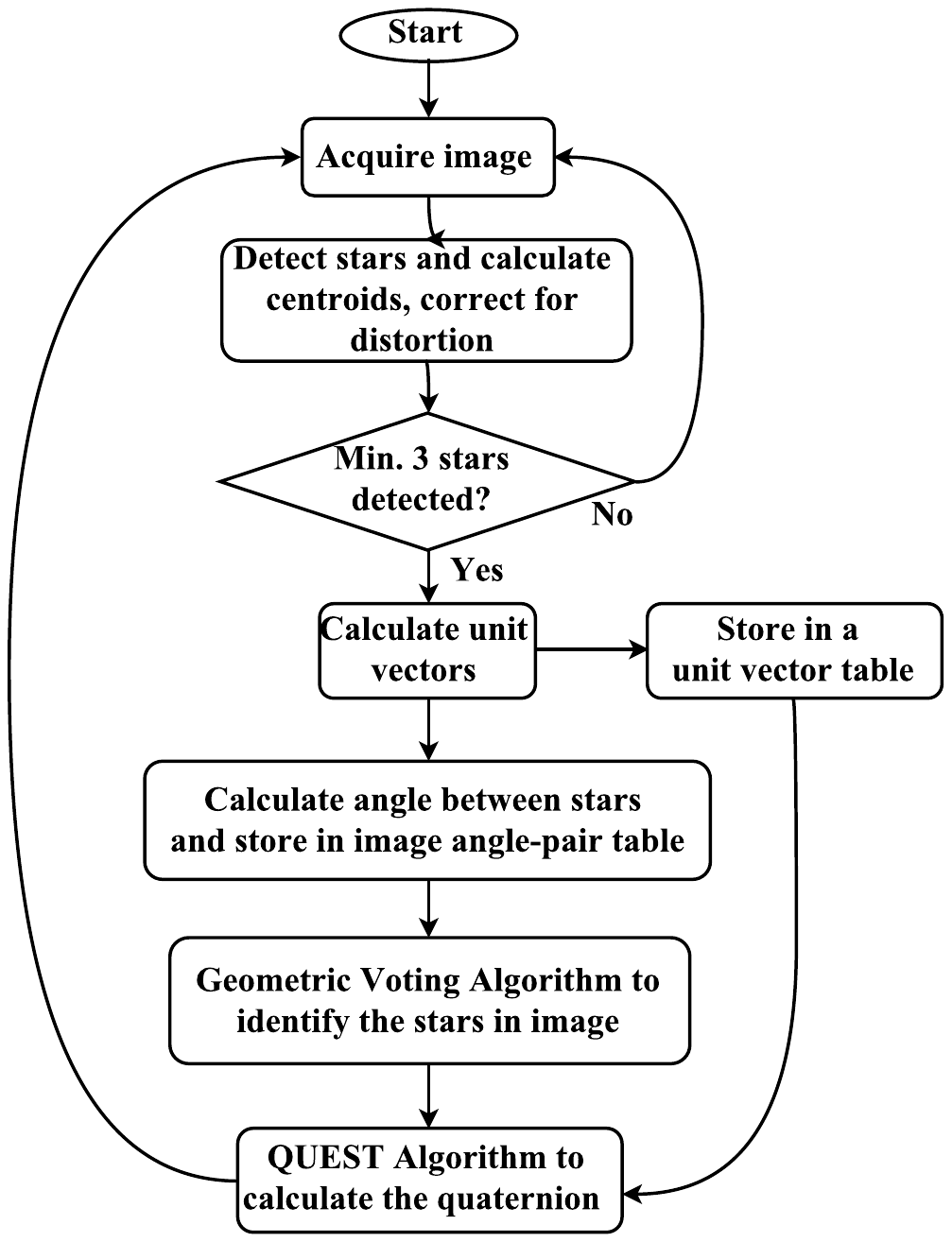}
\end{tabular}
\caption{A flowchart of the software architecture}
\label{fig:sw_gen_flowchart}
\end{figure}

All codes in this package are written in MATLAB \cite{matlab_software} due to its simplicity in scripting and excellent capabilities for visualizing the data. MATLAB (matrix laboratory) is a an environment and fourth-generation programming language for numerical computations and simulations developed by MathWorks\footnote{MATLAB is a licensed product which can be obtained from {\tt  https://in.mathworks.com/store}}. Fig.~\ref{fig:sw_gen_flowchart} depicts the flowchart of the software implementation. A Verilog module captures image from the detector and dumps it into the RAM provided on-board. After image acquisition, a microcontroller, programmed on the FPGA, starts a process of identification of regions corresponding to stars. Pixels having values higher than a predefined threshold are prospective stars. Stars are differentiated from other objects based on the number of pixels in a region. Centroids, which are essentially the ($x,y$) coordinates of the centers of the stars, are then calculated in the image plane. These are converted into unit vectors in the sensor body coordinate system through Eq.~(\ref{eq:centroid_2_unit_vectors}) below. The angles between stars are estimated by taking a dot product between every pair of stars visible in the image. Subsequently, the  Geometric Voting algorithm identifies the stars in the image by finding a match for the angle pairs from the angle pairs in the catalog. A binary search algorithm \cite{matlab_fex_bsearch} is used to search the angles from the catalog table. Lastly, the QUEST algorithm calculates the quaternion of the rotation between the ECI coordinate system and the sensor body coordinate system. 

\begin{figure}
\centering
\begin{tabular}{c}
\includegraphics[scale = 0.4]{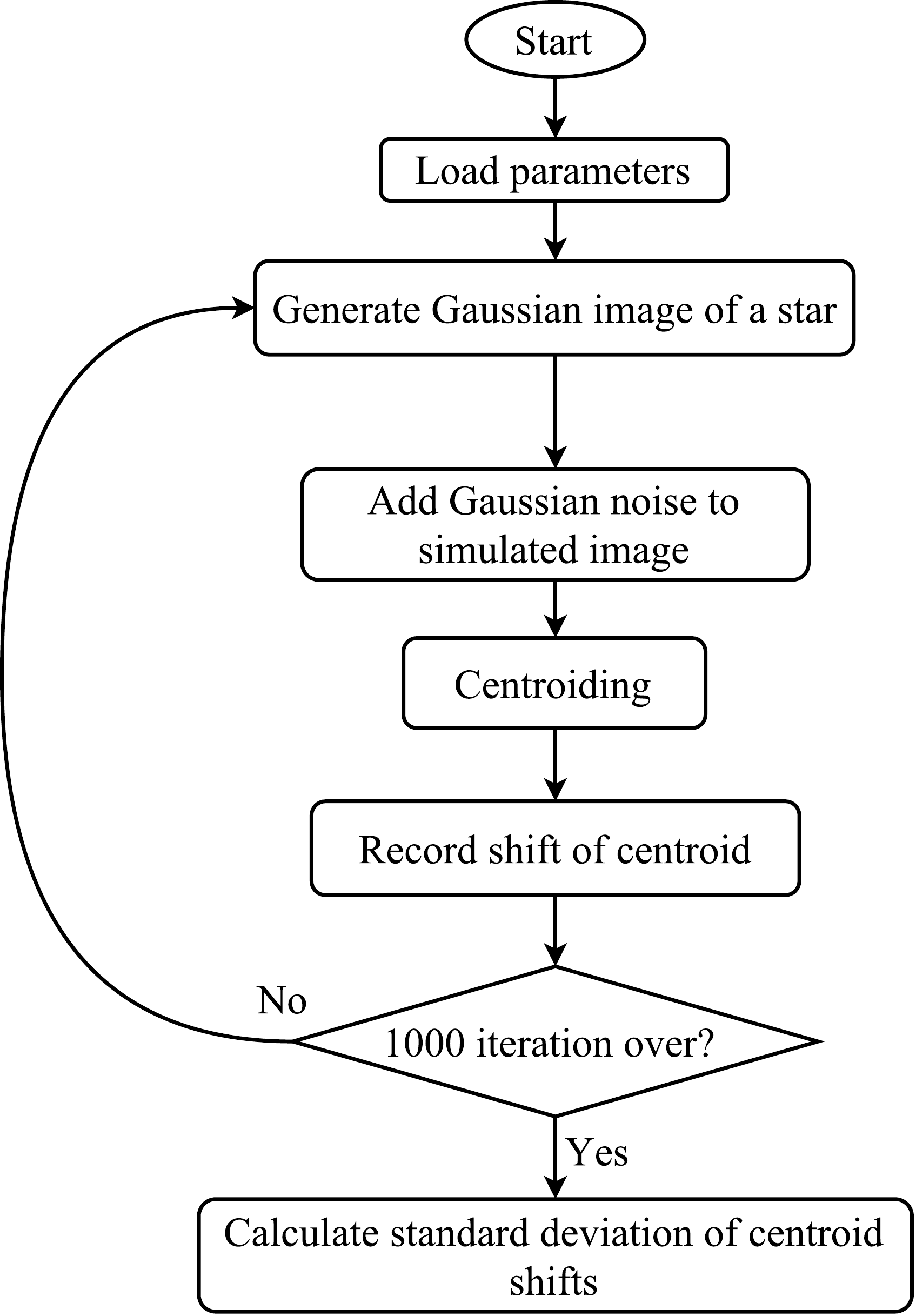}
\end{tabular}
\caption{Simulation of centroiding algorithm}
\label{fig:centroiding_sim_flowchart}
\end{figure}

\begin{figure}
\centering
\includegraphics[scale = 0.4]{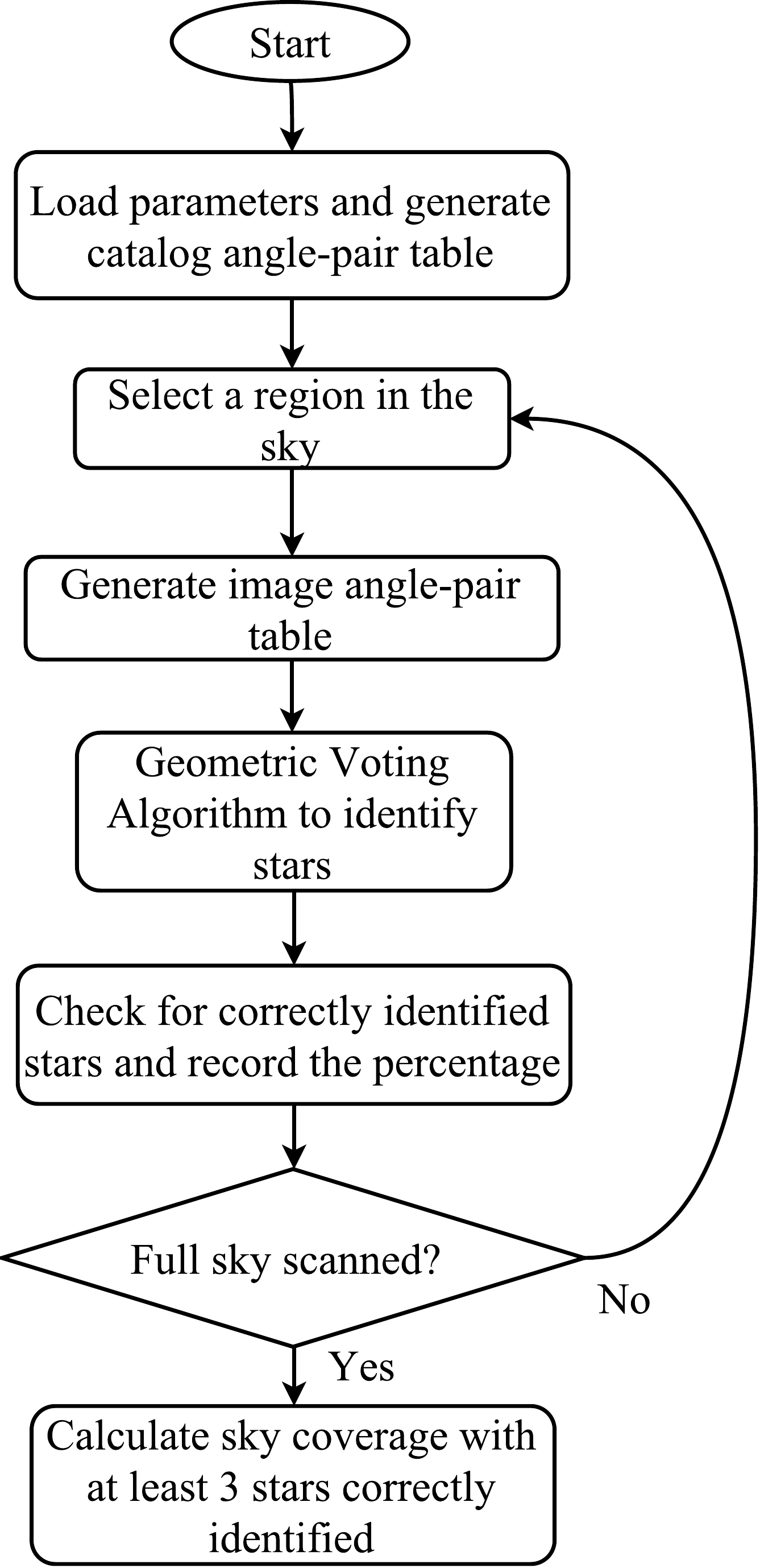}
\caption{Simulation of Geometric Voting algorithm}
\label{fig:gvalgo_sim_flowchart}
\end{figure}

\begin{figure}
\centering
\includegraphics[scale = 0.4]{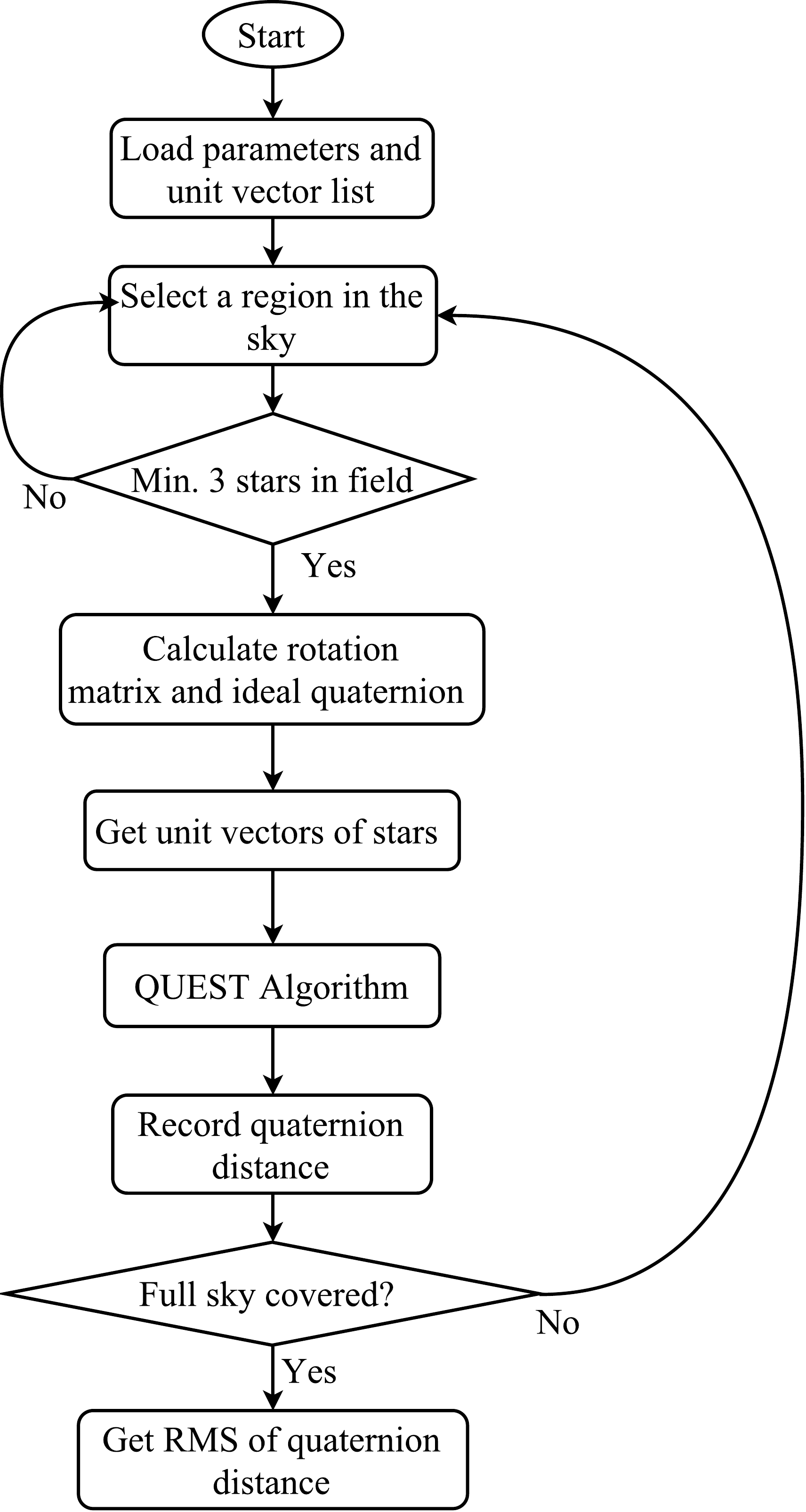}
\caption{Simulation of QUEST algorithm}
\label{fig:quest_sim_flowchart}
\end{figure}

To estimate the performance of the algorithms, the simulations were carried out in sequences  depicted in the following respective flowcharts:
\begin{itemize}
\item Centroiding algorithm -- Fig.~\ref{fig:centroiding_sim_flowchart}
\item Geometric voting algorithm -- Fig.~\ref{fig:gvalgo_sim_flowchart}
\item QUEST algorithm -- Fig.~\ref{fig:quest_sim_flowchart}
\end{itemize}
In the following sections, we describe the algorithms, the simulation of their performance and the results in detail.

\section{Algorithms}
\label{sec:algorithms}

\subsection{Coordinate Systems}

The procedure for obtaining attitude  quaternions from images needs clear definition of different coordinate systems. We use the following coordinate systems:
\begin{itemize}
\item{Image plane coordinate system}
\item{Sensor body coordinate system}
\item{Earth-Centered Inertial coordinate system}
\end{itemize}

The 2-dimensional image plane coordinate system describes a certain location in the detector plane. The origin of the coordinate system is the principal point, or the point of intersection of the optical axis with the image plane. The $x$ and $y$ axes are in the directions of the columns and rows of the image sensor area, respectively. The principal point of the optics is identified during the calibration of the camera. The coordinates are described in pixels or mm.  

The 3-dimensional sensor body coordinate system has its origin at the optical center. The $z$-axis is the direction of the lens boresight, and its $x$ and $y$ axis are aligned with the $x$ and $y$ axis of the image plane coordinate system. Sensor body coordinates are expressed in Cartesian form. Unit vectors of the stars in a given image are calculated in this coordinate system and are further used in Geometric Voting algorithm and QUEST algorithm. 

The Earth-Centered Inertial coordinate system is fixed in inertial space and thus used as the reference frame to calculate the quaternions. The $z$-axis points towards the North Celestial pole, and the $x$-axis is in the direction of vernal equinox. This is a unit-sphere coordinate system, and the position of any star is expressed in terms of only two angles: Right Ascension (RA), and Declination (Dec). The epoch in which the ECI coordinates of stars in the catalog are described, should be clearly mentioned, because the RA/DEC of stars change  over time due to precession, or proper motion.

\subsection{Centroiding Algorithm}

The image of a star on the detector is slightly defocussed and spread over multiple pixels. Therefore, the position of a star on the image plane is identified by calculating the centroid of the region where the star is imaged. 

The spread of a stellar image helps in filtering stars from erroneous hot pixels and extended objects. A region with very few bright pixels is a region of hot pixels or cosmic ray hit, and a region with more number of bright pixels than expected is an extended object. The optical design is made such that a star image is spread over a $4\times 4$ (16) pixel area. Therefore, any bright region which contains less than 10 pixels is rejected for being a hot pixel, and any bright region occupying more than 20 pixels is rejected for being an extended object. This filtering happens during the readout process and, thus, faulty star images do not propagate further in the calculation flow. Another advantage of spreading a star image is that the centroid coordinate can be calculated with sub-pixel accuracy, as opposed to the case when the centroid coordinates are integer numbers of the pixel position in $x$ and $y$ directions. 

The centroiding algorithm provides the centroid location in the image plane coordinate system. The computing procedure consists of three processes: 
\begin{itemize}
\item{Image plane search}
\item{Segmentation/region growing}
\item{Centroiding}
\end{itemize}

To identify a region as a star, we must first scan over the complete image area for pixels brighter than a pre-determined threshold value. This process is called the image plane search. The threshold value is identified during the calibration and real sky imaging. Since the size of the whole image is 1K$\times$1K, and each pixel is read from memory in approximately 10 ns, scanning of the complete image (1 million pixels) takes significant portion of the total time, from image acquisition to the  quaternion calculation. We check every alternate pixel to reduce the net time while still detecting stars successfully. 

On identifying a pixel with a value more than the pre-determined threshold, we start a process called region growing. In this process, we check 8 neighboring pixels for values greater than the threshold. On finding a neighboring pixel brighter than the threshold, we repeat the same region growing procedure for the neighboring pixel. During this looping routine, we also keep a count of the number of pixels in the region. Filtering of hot pixels/cosmic ray hits and extended objects happens at this stage. At the end of the loop, a region in the image which belongs to a star is identified. This region is allotted a number, and the image plane search is continued. At the end of the image plane search, all regions corresponding to stars are identified and given an image ID number.

After identifying all regions corresponding to stars in the image, we calculate the centroid of each region by weighing the average of that region,
\begin{align} \label{eq:centroid_x}
x &= \frac{\sum_{i=1}^{n} x_i \times I_i}{\sum_{i=1}^{n} I_i}\,,\\
\label{eq:centroid_y}
y &= \frac{\sum_{i=1}^{n} y_i \times I_i}{\sum_{i=1}^{n} I_i}\,,
\end{align}
where ($x,y$) are coordinates of the centroid, ($x_i,y_i$) are ($x,y$) coordinates of the $i^{th}$ pixel in the region, $I_i$ is intensity of the $i^{th}$ pixel in the region, and $n$ is the total number of pixels above the threshold value. The coordinates are described in the image plane coordinate system and are, thus, 2-dimensional. 

\subsection{Geometric Voting Algorithm}
\label{subsection:geometric_voting_algo}

Centroiding algorithm gives the coordinates of the stars in the image plane coordinate system. To identify  stars, we employ the Geometric Voting algorithm \cite{kolomenkin_paper}, where the angles between each pair of stars detected in the image are matched to the angles between pairs of stars in the star catalog. We convert the star positions in the image plane coordinate system to a unit vector in the sensor body coordinate system using Eq.~(\ref{eq:centroid_2_unit_vectors}), where ($u_x, u_y, u_z$) are components of the unit vector of the star in the sensor body coordinate system, ($x_u, y_u$) are coordinates of the centroids in the image plane coordinate system, ($x_c, y_c$) are coordinates of the principal point (i.e. the point where the optical axis intersects the detector) in the image plane coordinate system, ($pp_x, pp_y$) are pixel sizes in $x$ and $y$ directions of the image sensor, respectively, and $f_{mm}$ is the focal length of the imaging optics.   

\begin{strip}
\begin{equation}
\begin{bmatrix} u_{x} \\ u_{y} \\ u_{z}\end{bmatrix} = \left(1 + \left(\left(x_{u} - x_{c}\right)\frac{pp_x}{f_{mm}}\right)^{2} + \left(\left(y_{u} - y_{c}\right)\frac{pp_y}{f_{mm}}\right)^{2} \right) ^{-\frac{1}{2}}
\begin{bmatrix}
(x_{u} - x_{c})\frac{pp_x}{f_{mm}}\\
(y_{u} - y_{c})\frac{pp_y}{f_{mm}}\\
1
\end{bmatrix}\,,
\label{eq:centroid_2_unit_vectors}
\end{equation}
\end{strip}

After conversion to unit vectors in sensor body coordinate system, the angle between each pair of imaged stars is obtained by taking dot products of their unit vectors. The pairs of image ID numbers and the angle between them are tabulated in an image angle-pair table. 

A star catalog is divided into two parts: a unit vector list and a catalog angle-pair table. These lists/tables are generated using Hipparcos catalog of nearby stars as a base\footnote{available for download from http://www.heasarc.gsfc.nasa.gov} by selecting stars brighter than a certain limiting magnitude and calculating their unit vectors in the ECI coordinate system. The catalog ID of the star (HID), along with its unit vector, is tabulated in unit vector list. 
A catalog angle-pair table is generated by taking dot products of pairs of unit vectors, and HIDs of the pair of stars with their angular distance are tabulated. We keep only the entries that are smaller than the FOV of the optics. We sort the catalog angle-pair table in increasing order of the angle values. Both parts of the catalog are stored in the on-board flash memory (Table~\ref{table:star_sensor_electronics}), and are loaded into the RAM upon initializing the star sensor.  

The Geometric Voting algorithm finds a match between each entry in the image angle-pair table with the catalog angle-pair table using a binary search method. All the entries in the catalog angle-pair table with angle values lying in a small range around the angle value in the image angle-pair table are selected. The imaged stars can possibly be any pair of stars from this selected star pairs. The HIDs of stars in the catalog angle-pair table cast a vote for the identified stars. This process goes on for all the entries in the image angle-pair table. At the end we find the maximum number of votes given for each imaged star. The HID corresponding to maximum number of votes is the most probable match. Further we start a verification process where the angle between each pair of stars with HIDs is verified to be lying in a small range around the angle between corresponding pair of stars in the image. This process verifies whether the identification of stars was done correctly.

\subsection{Quaternion Estimator (QUEST) Algorithm}

After identification of stars in the image, the last step is to calculate the quaternion of rotation between the ECI coordinate system and the sensor body coordinate system. This process is called attitude determination and can be implemented using various algorithms \cite{attitude_estimate_from_vector_obs,shuster_paper_QUEST_for_better_attitudes,shuster_basic_QUEST_paper}.
In our case, the output of the Geometric Voting algorithm gives the unit vectors of the stars in ECI coordinate system, and unit vectors in the sensor coordinate system are calculated from Eq.~(\ref{eq:centroid_2_unit_vectors}). These two set of unit vectors are inputs to the QUEST algorithm. 

The rotation between any two coordinate systems can be described in different ways, such as rotation matrix, Euler angles, quaternions, etc. Each of these descriptions have their own applications and their own pros and cons. We select quaternions to describe the rotation/orientation,  because this method is comparatively less computationally extensive. 

The main aim of an attitude estimation algorithm is to use a pair of vector lists, one in ECI coordinate system and the other in sensor body coordinate system, to determine their relative rotation. 

Let the unit vectors in ECI coordinate system be $\mathbf{v}_i$, and those in the sensor body coordinate system be $\mathbf{v}_b$. The equation $\mathbf{v}_b=R_{bi}\mathbf{v}_i$ holds true in an ideal case, where $R_{bi}$ defines the rotation matrix between the ECI coordinate system and the sensor body coordinate system. Our problem is to identify $R_{bi}$ from the unit vectors list obtained in the previous steps. One approach is a statistical approach, where we get many measurements of the $\mathbf{v}_b$ from the sensor body coordinate system in the form of star locations and identify those stars by Geometric Voting algorithm which gives $\mathbf{v}_i$ in the ECI coordinate system. We need to find $R_{bi}$ which minimizes the loss function,
\begin{equation}
J(R_{bi}) = \frac{1}{2} \sum_{k=1}^{n} \omega_{k}\left| \mathbf{v}_{kb} - R_{bi}\mathbf{v}_{ki}  \right|^{2}\,,
\label{eq:loss_function}
\end{equation}
where $J$ is the loss function to be minimized, $\omega_k$ is the set of weights assigned to each vector pair measurement, and $n$ is the number of correctly identified stars. This problem is called the Wahba's problem \cite{Wahba} and different solutions for this problem are suggested in the form of different methods to calculate the attitude information. In an ideal case when all measurements are perfect, we obtain $J = 0$. However in practice, there will always be an error in measurement resulting in $J > 0$. The smaller $J$ can be made, the better is the approximation of $R_{bi}$. We restate the loss function in terms of quaternions in such a way that it becomes an eigenvalue problem, where the largest eigenvalue is to be found. However, finding an eigenvalue is computationally very intensive for an embedded system. The QUEST algorithm was developed to bypass the expensive eigenvalue problem by approximating this process  \cite{attitude_estimate_from_vector_obs}. The final form of the problem involves solving for $\mathbf{p}$ in the following equation
\begin{equation}
\left[ \left( \lambda_{opt} + \sigma \right) I - S \right] \mathbf{p} = Z\,,
\label{eq:rodriguez_parameter}
\end{equation}
where 
\begin{align*}
&B = \sum_{k=1}^{n}\omega _{k}\left ( v_{kb}v_{ki}^{T} \right) \,,\\
&\lambda_{opt} = \sum\omega_k \,,\\
&\sigma = \mbox{Tr($B$), i.e. sum of diagonal elements of $B$}\,,\\
&S = B + B^T\,,\\
&\mathbf{p} \,\mbox{is the Rodriguez parameter which must be solved for}\,,\\
&Z = \left[ \begin{matrix}
B_{23}- B_{32} & B_{31}-B_{13} & B_{12}- B_{21}
\end{matrix} \right]^{T}\,.
\end{align*}
Once the Rodriguez parameter $p$ has been found, the attitude quaternion can be calculated using 
\begin{equation}
{\mathbf{q}_{quest}} = \frac{1}{\sqrt{1+\mathbf{p^{T}}\mathbf{p}}}\left [ \begin{matrix}
\mathbf{p}\\ 
1
\end{matrix} \right ]\,,
\label{eq:rodrigueq2quaternion}
\end{equation}
where $\mathbf{q}_{quest} = \left [ \begin{matrix}
\mathbf{q_{1}}\\ 
\mathbf{q_{2}}\\ 
\mathbf{q_{3}}\\ 
q_{4}
\end{matrix} \right ] = q_1\mathbf{i} + q_2\mathbf{j} + q_3\mathbf{k} + q_4$.

\section{Simulations, Performance and Results}
\label{sec:performance_est}

\subsection{Centroiding Algorithm}

To simulate every step in this algorithm, we require an image of a star (reference image) with characteristics of the star sensor optics. We simulate this image by generating a Gaussian PSF with FWHM of 2 pixels (Sec.~\ref{sec:hardware_info}). This image is generated by a 2-dimensional function for a $1024\times 1024$ pixel array, 
\begin{align}  
\label{eq:gaussian_func}
G(x,y) &= \frac{1}{\sigma_{\rss G}\sqrt{2\pi}}\exp^{-\frac{(x-x_{c})^2 + (y-y_{c})^2}{2\sigma_{\rss G}^2}}\,,\\
\label{eq:fwhm_sigma}
\sigma_{\rss G} &= \frac{FWHM}{2\sqrt{2\log2}}\,,
\end{align}
where:
\begin{trivlist}
\item $x$ varies as $1,2,\dots 1024$;
\item $y$ varies as $1,2,\dots 1024$;
\item $(x_c,y_c)$ is the coordinate of the ideal center of the star in the image plane coordinate;
\item $\sigma_{\rss G}$ is the variance of the Gaussian function;
\item FWHM is the FWHM of the star image as expected from the star sensor lens.
\end{trivlist}

Because the exposure time of the image is always the same, i.e. 100 ms, we can estimate the noise introduced by the readout mechanism, dark noise and fixed pattern noise. Knowing these characteristics and assuming various signal to noise ratio (SNR) for the star image, we have added Gaussian noise, matching the pixel noise in the  image sensor, to the star image. The mean value of the Gaussian noise is the value of the dark signal, and its standard deviation $\sigma$ is the sum of the readout and the dark noise. 

\subsubsection{Processing Time}

A significant fraction of the time of the centroiding algorithm is utilized in the image plane search process, and to reduce the time required in this process, we search only every alternate pixel. The time required to scan through the complete image reduces to almost $1/4^{th}$ of the default case. Sampling at every $4^{th}$ pixel speeds up the process even further, but we might miss the actual star because FWHM=2 pixel, by default. Times taken for each search methods are tabulated in Table~\ref{table:time_simulation}. Simulations performed with different SNR suggest that we do not miss any star with SNR $> 3\,\sigma$ if we use the alternate pixel scanning.

\begin{table} 
\centering
\caption{Times recorded for image plane search (as simulated in MATLAB).}
\begin{tabular}{ccc}
\hline
Pixel increment & Number of pixels checked & Time required\\
\hline
1 pixel & 1048576 & 0.0956 sec\\
2 pixel & 262144 & 0.0293 sec \\
4 pixel & 65536	& 0.0102 sec \\
\hline
\end{tabular}
\label{table:time_simulation}
\end{table}

\subsubsection{Centroiding Accuracy}
\label{subsection:centroiding_accuracy}

We simulated an implementation of the complete centroiding algorithm in a sequence including image plane search, region growing/segmentation, and finally centroiding. This is performed on a reference image where, unlike in the ideal case, the calculated centroid has positional uncertainty due to added noise. The simulation is repeated 1000 times, each time generating a randomly varying Gaussian noise at each pixel, and the shift in the centroid is recorded for each image. The mean shift in these 1000 images gives a measure of the uncertainty of the centroiding algorithm due to the introduced noise. This uncertainty varies with the star magnitude and SNR as shown in Table~\ref{table:snr_accuracy_simulation}.

\begin{table}[h!]
\centering
\caption{ Variation of centroiding error with SNR of source}
\begin{tabular}{ccc}
\hline
Magnitude & SNR & Centroid shift \\
\hline
0 & 311.98 & 	1.2 	\\
0.5 & 247.78 & 	0.80	\\
1 & 196.78 & 	1.13	\\
1.5 & 156.26 & 	1.53	\\
2 & 124.06 & 	1.83 	\\
2.5 & 98.46 & 	2.03	\\
3 & 78.11 & 	3.50	\\
3.5 & 61.92 & 	3.67	\\
4 & 49.04 & 	5.85	\\
4.5 & 38.75 & 	16.35	\\
5 & 30.55  	&  	23.25	\\
5.5 & 23.97	&	27.68	\\
6 & 18.69	&	34.35	\\
6.5 & 14.44 &	39.6	\\
\hline
\end{tabular}
\label{table:snr_accuracy_simulation}
\end{table}

\subsubsection{Memory requirement}
The centroiding process occupies the greatest part of the total temporary storage memory (random access memory -- RAM). All subsequent processing requires only the storage of the centroid coordinates during one iteration. We estimated the RAM requirements of the centroiding algorithm through basic calculations. The image sensor has a size of $1024 \times 1024$ pixels, with 10 bits per pixel. Thus, the  memory requirement for one image is $1024\times 1024 \times 10$ bits $\approx 1$ MB. Apart from this, we have to store an image mask, which identifies the regions of the image that belong to a star, which accounts for another 1 MB of RAM. Furthermore, the image acquisition and image processing are run as two parallel tasks, which means that we need another 2 MB of RAM for the implementation. So much of RAM (4 MB) is not available on the FPGA chip and, therefore, we added an external RAM chip to the image processor PCB (Table~\ref{table:star_sensor_electronics}). 

\subsection{Geometric Voting Algorithm}
The Geometric Voting algorithm is basically the implementation of a voting scheme. The inputs for this algorithm are the catalog angle--pair table and the image angle--pair table. The algorithm outputs the HID of the stars in the image. To evaluate the performance of the algorithm in terms of sky coverage and identification/verification accuracy, we simulate the ideal inputs required for the algorithm (i.e. the image angle--pair table) using physical parameters of the optics (FOV and limiting magnitude) and selecting sources from the Hipparcos star catalog. 

\begin{figure*}
\begin{center}
\begin{tabular}{c}
\hspace{-0.5in}
\includegraphics[scale=0.55]{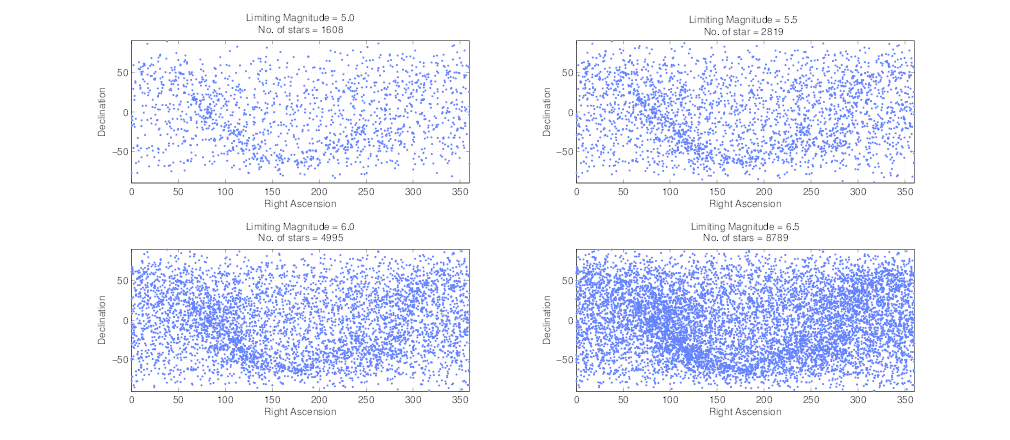}
\end{tabular}
\end{center}
\caption{Star distribution at different limiting magnitudes}
\label{fig:different_limiting_mag_skies}
\end{figure*}

\subsubsection{Sky Coverage}

Here we verify the performance of the Geometric Voting algorithm in correctly identifying stars in any given field in the sky. 
We know that the FOV of the optics is $10\degree$, and that its limiting magnitude is $6.5^m$. By varying the threshold in the centroiding procedure, we can adjust limiting magnitudes of detected stars in the image. We consider various limiting magnitudes from 5.0 to 6.5 in the analysis. The distribution of stars in the sky with different limiting magnitudes is shown in Fig.~\ref{fig:different_limiting_mag_skies}.
The catalog angle--pair table is formed using these stars. Different limiting magnitude skies have different number of entries in the catalog angle--pair table depending on the number of stars in the sky (Table~\ref{table:entries_catalog_angle_pair_table}). The size of this catalog is taken into consideration while deciding on the flash memory size (Table~\ref{table:star_sensor_electronics}).

\begin{table}
\centering
\caption{Number of entries in catalog angle--pair table}
\begin{tabular}{ccc}
\hline
Limiting Magnitude & No. of stars & No. of entries \\
\hline
5.0 & 1608 	& 11671 \\
5.5 & 2819	& 35493 \\
6.0 & 4995	& 108656 \\
6.5 & 8789	& 332092\\
\hline
\end{tabular}
\label{table:entries_catalog_angle_pair_table}
\end{table}

\begin{table}
\centering
\caption{Minimum and maximum number of stars in any field}
\begin{tabular}{ccc}
\hline
Limiting  & Minimum no.  & Maximum no.  \\
Magnitude & of stars & of stars \\
\hline
5.0 & 1 & 18 \\
5.5 & 1	& 27 \\
6.0 & 1	& 39\\
6.5 & 4	& 70\\
\hline
\end{tabular}
\label{table:max_min_stars}
\end{table}

\begin{figure*}
\begin{center}
\begin{tabular}{cc}
\includegraphics[scale=0.4]{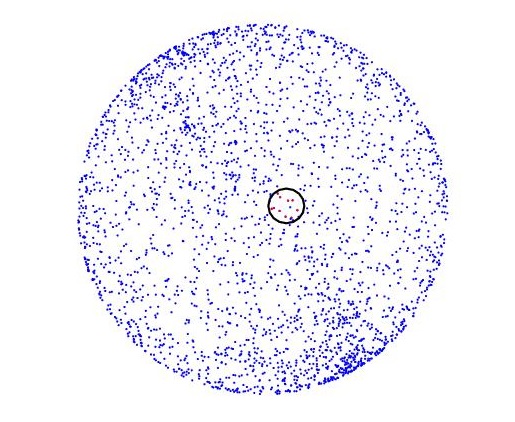} & 
\includegraphics[trim=175 10 100 10,clip,scale=0.5]{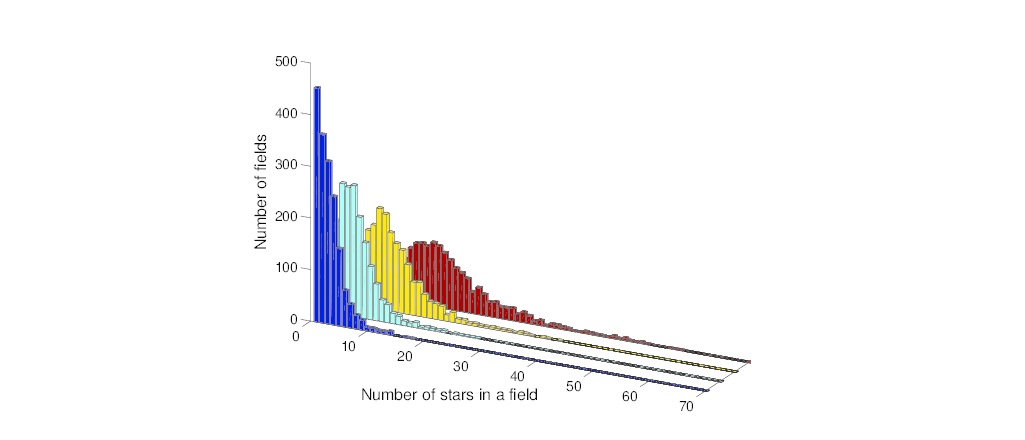} 
\end{tabular}
\end{center}
\caption{{\it Left}: Selecting a field in the sky matching with camera FOV (10$\degree$). {\it Right}: Histogram of the number of fields vs. the number of stars in a field. Blue, cyan, yellow and red are for limiting magnitudes $5.0$, $5.5$, $6.0$ and $6.5$, respectively.}
\label{fig:selecting_field_histogram}
\end{figure*}

We divide the entire sky into 1728 overlapping fields, where each field is $10\deg$ (optics FOV) and fields are spaced at $5^{\circ}$ from each other (Fig.~\ref{fig:selecting_field_histogram}, {\it Left}). From Hipparcos catalog, we find the number of stars in each field for different limiting magnitudes (Fig.~\ref{fig:selecting_field_histogram}, {\it Right}). The minimum and maximum number of stars found in any field are tabulated (Table~\ref{table:max_min_stars}). For example, for a limiting magnitude of $5.0$, the minimum number of stars in any field is $1$. This shows that at any pointing of the telescope, there is always at least $1$ star in the FOV. 

Using the unit vectors of the stars in the selected field, we generate an image--angle pair table. The stars are renumbered in the order of their appearance in the selected field, keeping their HIDs in a separate variable. The Geometric Voting algorithm estimates the HID of a star in the selected field and verifies it by matching its angle in the image--angle pair table with the catalog--angle pair table. This process is repeated for every field. In Section~\ref{subsection:centroiding_accuracy}, we have estimated centroiding accuracy of the optics for different magnitudes and corresponding SNR values, where the worst centroiding accuracy is $\sim35\as$ for $6^m$ stars. This is the range within which the Geometric Voting algorithm looks for the matches in catalog--angle pair table. We introduce the centroiding error arising in measuring the angles between the imaged stars, into the image--angle pair table, to simulate a realistic situation. We use the normal distribution with mean value of the worst-case of centroiding accuracy ($35\as$) to simulate the noise to be added. The results obtained from this analysis are shown in Table~\ref{table:gva_sky_coverage}.

\begin{table}[h!]
\centering
\caption{Results of Geometric Voting Algorithm Sky Coverage}
\begin{tabular}{lcccc}
\hline
                                                  & \multicolumn{4}{c}{Limiting Magnitudes} \\
                                                  & 5.0     & 5.5      & 6.0      & 6.5     \\
\hline
No. of fields    &  910	& 1437	& 1704	& 1728 \\
Correctly estimated & 37.14\%    & 64.02\%     & 88.43\%     & 98.95\%    \\
Correctly verified  & 37.14\%    & 64.02\%      & 88.43\%    & 98.95\%  \\
\hline
\end{tabular}
\label{table:gva_sky_coverage}
\end{table}

\subsubsection{Timing Analysis}

\begin{figure}
\centering
\includegraphics[trim=275 10 150 10,clip, scale = 0.5]{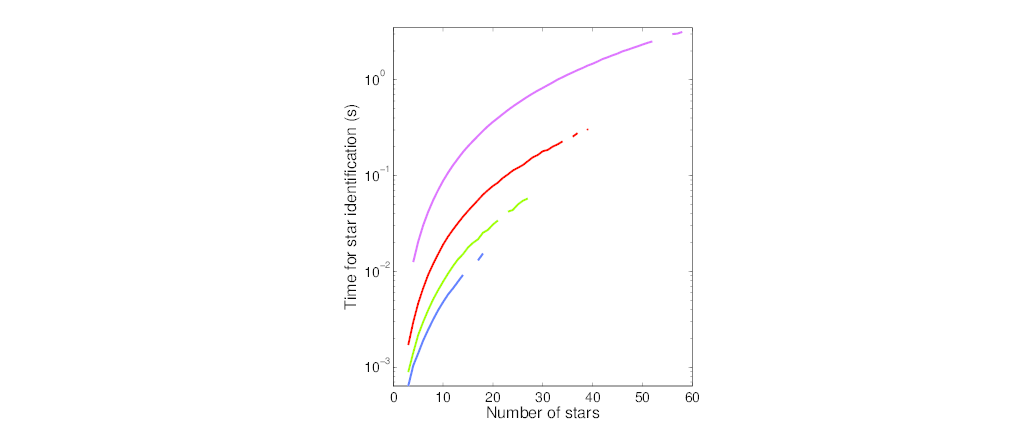} 
\caption{Time required for identification with varying number of stars in field. Blue, green, red and magenta lines show the timings for limiting magnitudes 5.0, 5.5, 6.0 and 6.5, respectively}
\label{fig:timing_across_magnitudes}
\end{figure}

During the procedure of estimating the sky coverage, we also estimate the time required to identify the stars in the field. This time varies with the number of stars visible in the selected field. It is also affected by the catalog angle-pair table selected depending upon the limiting magnitude, because the number of entries in catalog angle-pair table is different for different limiting magnitudes (Table~\ref{table:entries_catalog_angle_pair_table}). The required times are shown in Fig.~\ref{fig:timing_across_magnitudes}. Blue, green, red and magenta lines show the times for limiting magnitudes  5.0, 5.5, 6.0 and 6.5, respectively. It can be seen that as the number of stars in a certain field increases, the time required to identify them also increases.

\subsection{Quaternion Estimator (QUEST) Algorithm}

The QUEST algorithm takes the unit vectors of reference points in the ECI and sensor body coordinate systems as inputs, and yields the estimated quaternions of rotation between the coordinate systems as output. To estimate the performance of this algorithm, we simulate the ideal inputs in a similar way as we did for the Geometric Voting algorithm. We already know the unit vectors of stars in ECI coordinate system visible in each field from the unit vector list of the star catalog as described in Section~\ref{subsection:geometric_voting_algo}. We simulate their unit vectors in sensor body coordinate system by the following process. The rotation matrix between the ECI coordinates and the sensor body coordinates for a selected field is calculated from the following equation,
\begin{equation}
R_{bi} = \left [ \begin{matrix}
\cos\alpha \cos\delta  &-\cos\alpha \sin\delta   &\sin\alpha  \\ 
\sin\delta & \cos\delta   & 0 \\ 
-\sin\alpha \cos\delta & \sin\alpha \sin\delta  & \cos\alpha  
\end{matrix} \right ]
\label{eq:rotation_matrix_particular_field}
\end{equation}
where $\alpha$ and $\delta$ are the RA and Dec of the center of the field the camera is looking at. The unit vectors in sensor body coordinate system of the stars visible in that field are the product of the rotation matrix with their unit vectors in the ECI coordinate system,
\begin{equation}
\mathbf{v}_b = R_{bi}\mathbf{v}_i\,,
\label{eq:vectors_rotation_matrix}
\end{equation}
where $\mathbf{v}_b$ is the unit vector of a particular star in the sensor body coordinate system, and $\mathbf{v}_i$ is the unit vector of that same star in the ECI coordinate system. The ideal quaternion is obtained from $R_{bi}$ (Eq.~\ref{eq:rotation_matrix_particular_field}) by
\begin{equation}
\mathbf{q}_{ideal} = \left [ \begin{matrix}
\mathbf{q}_1\\ 
\mathbf{q}_2\\ 
\mathbf{q}_3\\ 
q_4
\end{matrix} \right ] =  
\begin{bmatrix}
\dfrac{m_{21} - m_{12}}{2\sqrt{1+m_{00}+m_{11}+m_{22}}}\\ 
\\
\dfrac{m_{02} - m_{20}}{2\sqrt{1+m_{00}+m_{11}+m_{22}}}\\ 
\\
\dfrac{m_{10} - m_{01}}{2\sqrt{1+m_{00}+m_{11}+m_{22}}}\\ 
\\
\dfrac{\sqrt{1+m_{00}+m_{11}+m_{22}}}{2}
\end{bmatrix} \,,
\label{eq:rotation_matrix2quaternions}
\end{equation}
where $m_{00}$, $m_{01}$, $m_{02}$, $m_{10}$,... $m_{22}$ are the elements of the rotation matrix. 
The output quaternion from the algorithm is then compared with this quaternion by calculating the distance between these quaternions using their dot product 

\begin{equation}
d = cos^{-1}(2\mathbf{q}_{ideal}\mathbf{.}\mathbf{q}_{quest} - 1)
\label{eq:difference_between_quat}
\end{equation} 

\subsubsection{Sky Coverage and Accuracy}

For quaternion calculation we have selected a criterion of having minimum of 3 stars in any given field. Therefore, the number of fields in the sky with more than 3 stars in a field defines the sky coverage for the QUEST algorithm. The quaternion distance (Eq.~\ref{eq:difference_between_quat}) is calculated for all the fields in the sky. Its standard deviation (RMS error) gives the accuracy of the QUEST algorithm, which is almost independent on limiting magnitude and very small, of the order $\sim 10^{-7}$ deg. The number of fields in the sky with more than 3 stars in a field are tabulated in Table~\ref{table:sky_coverage}.

\begin{table}[h!]
\centering\caption{Performance of QUEST for different limiting magnitudes}
\begin{tabular}{cc}
\hline
Limiting Magnitude & Sky Coverage \\
\hline
5.0 & 52.66\% \\
5.5 & 83.16\% \\
6.0 & 98.61\% \\
6.5 & 100\% \\
\hline
\end{tabular}
\label{table:sky_coverage}
\end{table}

\section{Conclusion}

\begin{table}
\centering
\begin{tabular}{lc}
\hline
Parameter & Value \\
\hline
Centroiding accuracy	&	$35\as$ \\
Distortion residual		&	$8\as$ \\
Centroiding time 		& 	0.05 sec \\
QUEST numerical error	&	$\sim 10^{-7\prime \prime}$ \\
Realtime memory requirement &	$\sim 4 MB$ \\
Catalog size 			&	$\sim 1.5 MB$ \\
Sky Coverage			&	$90\%$ \\
\hline
\end{tabular}
\caption{Estimated parameters from the software package}
\label{table:budget}
\end{table}

We have put together a complete software package to reduce star sensor images to quaternions and to evaluate the performance of the operational algorithms. The package simulates centroiding, Geometric Voting and QUEST algorithms, and evaluates such performance parameters as attitude accuracy, calculation time, required memory, star catalog size, sky coverage, etc., and estimated the errors introduced by each algorithm. The testing is parametrized for different hardware parameters of the star sensor, such as the focal length of the imaging setup, FOV of the camera, angle measurement accuracy, distortion effects, and others. We conclude with the following remarks: 
\begin{itemize}
\item For our {\em StarSense}, we find that a limiting magnitude of V=6.0 is optimal to get significant sky coverage with minimal calculation time. 
\item This software package is robust, fast, user-friendly in terms varying hardware parameters, and easily portable to various operating  platforms. 
\item Due to the parametrized approach in  package development, it can be applied to evaluate the performance of such algorithms in any star sensor. 
\end{itemize}
The estimated performance parameters and the estimated errors are tabulated in Table~\ref{table:budget}. The source codes of the software package can be obtained from a github repository \cite{github_source}.

\section{Acknowledgments}
Part of this research has been supported by the Department of Science and Technology (Government of India) under Grant IR/S2/PU-006/2012.

\end{document}